\documentclass[10pt]{amsart}
\usepackage[all]{xy}
\usepackage{amsmath}
\usepackage{amsfonts}
\usepackage{amssymb}
\usepackage{amscd}
\usepackage{amsthm}
\usepackage{latexsym}
\usepackage{amsbsy}
\usepackage{graphicx}
\usepackage{epstopdf}
\newtheorem{lemma}{Lemma}[section]

\newtheorem{theorem}{Theorem}[section]
\newtheorem{corollary}{Corollary}[section]

\begin{document}

\title{Spectral Action for Bianchi Type-IX Cosmological Models}

\author{Wentao Fan, Farzad Fathizadeh, Matilde Marcolli}

\thanks{E-mails: wffan@caltech.edu, farzadf@caltech.edu, matilde@caltech.edu}

\maketitle

\begin{center}
Division of Physics, Mathematics and Astronomy \\ 
California Institute of Technology \\
Pasadena, California, USA
\end{center}

\begin{abstract}

A rationality result previously proved for 
Robertson-Walker metrics is extended to 
a homogeneous anisotropic cosmological model, 
namely the Bianchi type-IX minisuperspace. 
It is shown that the Seeley-de Witt coefficients 
appearing in the expansion of the spectral action for the 
Bianchi type-IX geometry are 
expressed in terms of polynomials with rational 
coefficients in the cosmic evolution 
factors $w_1(t), w_2(t), w_3(t),$ and their higher derivates 
with respect to time.  
We begin with the computation of the Dirac operator 
of this geometry and calculate the coefficients $a_0, a_2, a_4$ 
of the spectral action by using heat kernel methods and parametric 
pseudodifferential calculus. An efficient method is devised for 
computing the Seeley-de Witt coefficients of a geometry 
by making use of  Wodzicki's noncommutative residue, 
and it is confirmed that the method checks out 
for the cosmological model studied in this article. 
The advantages of the new method are discussed, 
which combined with symmetries of the Bianchi type-IX 
metric, yield an elegant proof of the rationality result.

\end{abstract}

\date{}



\tableofcontents

\section{Introduction}

Quantum cosmology studies the early universe where the energy scale
is so high that one would need to incorporate into the theory both
quantum gravity and the Standard Model with unbroken symmetries 
such as the hypothetical supersymmetry. This makes the exact solution of
any quantum cosmological problem essentially impossible as it amounts to 
solving quantum fields genuinely interacting with quantized gravity,
the nature of which we know little about. However, since we are only
interested in the large-scale behavior of the early universe whose
mass-energy distribution is highly homogeneous, we are encouraged 
to exploit this symmetry and focus only on a few long-wavelength degrees
of freedom. In quantum cosmology, this common practice is known as
the minisuperspace approximation, which can be rigorously justified
under certain criteria \cite{KucRya, SinHu}.

Well-known examples of minisuperspaces include the Robertson-Walker 
model and its anisotropic generalization to the Bianchi type-IX model. 
The  Robertson-Walker metric is of the form 
\[
ds^2 = dt^2 + a(t)^2 d\sigma^2, 
\] 
where $a(t)$ is a general cosmic factor of the expanding universe and $d \sigma^2$ 
is the round metric on the 3-sphere $\mathbb{S}^3$. The Bianchi type-IX model, which 
enjoys a reduced $SU(2)$ isometry group rather than the full $\mathbb{S}^{3}$ 
symmetry, is written as 
\begin{equation} \label{Bianchimetric}
ds^2 = w_1(t) w_2(t) w_3(t) dt^2 + 
\frac{w_2(t) w_3(t)}{w_1(t)} \sigma_1^2 + 
\frac{w_3(t) w_1(t)}{w_2(t)} \sigma_2^2+ 
\frac{w_1(t) w_2(t)}{w_3(t)} \sigma_3^2,    
\end{equation}
where $\sigma_i$ are left-invariant 1-forms on $SU(2)$-orbits.

Beside trying to fit the Standard model into the early universe spacetime,
another possible investigation of physical importance is to quantize
supersymmetric systems in a Bianchi-type IX minisuperspace. Since
the energy scale of the early universe is very likely to be higher
than the supersymmetry breaking scale, it is of great interest  to see how 
supersymmetry may possibly change the picture
of the early universe. Since the Hamiltonians of supersymmetric systems
can often be identified with well-behaving elliptic operators acting
on spin bundles for Majorana fermions or fields of differential
forms for Dirac fermions, the {\it spectral action} \cite{ChaConSAP} approach appears to
be a handy tool, where the celebrated Atiyah-Singer index theorem
may be used to calculate the Witten indices of certain supersymmetric
theories and thus to conclude whether the supersymmetry can be spontaneously
broken \cite{Wit, Alv}.

Noncommutative geometry \cite{ConBook} describes geometric spaces 
by spectral triples $(\mathcal{A},\mathcal{H}, D)$,
where $\mathcal{A}$ is an involutive algebra represented by bounded operators on a 
Hilbert space $\mathcal{H}$, and $D$ is an unbounded self-adjoint operator in 
$\mathcal{H}$ that plays the role of the Dirac operator by encoding the metric information. 
This set up, which includes a great variety of noncommutative spaces,  generalizes  
Riemannian geometry since Connes' reconstruction theorem \cite{ConReconstruct} states 
that if $\mathcal{A}$ is commutative, then, under suitable regularity conditions, the triple 
consists of the algebra of smooth function on a spin$^c$ manifold acting on the $L^2$-spinors 
and $D$ is the Dirac operator.

Given a spectral triple, the spectral action principle \cite{ChaConSAP, ConMarBook} 
considers the functional,  
\[
\text{Trace} \big ( f (D/ \Lambda)\big ) \sim  \sum_{\beta \in \Pi} 
f_\beta \Lambda^\beta \int\!\!\!\!\!\!- \, |D|^{-\beta}+ f(0) \zeta_D(0) + \cdots, 
\]
as the fundamental action functional, where $f$ is a positive even function on the real 
line and  $\Lambda$ is a real parameter that fixes the mass scale.  The details of the 
above asymptotic expansion are explained in \cite{ConMarBook}. For  a spin$^c$ manifold $M$, 
the coefficients of this expansion are determined by the Seeley-de Witt coefficients $a_n$ 
appearing in the heat expansion, 
\[
\text{Trace}\big( e^{-t D^2}   \big ) 
\sim 
t^{- \text{dim} (M)/2} \sum_{n=0}^\infty a_{2n}(D^2) t^n \qquad ({t \to 0^+}). 
\] 

In addition to extracting local geometric information
and recovering the Einstein-Hilbert action, the spectral action  generalizes the 
theory to include noncommutative spaces which give rise to the Standard Model
gauge fields with Yang-Mills actions as well as couplings to the modified
gravity \cite{ChaConUncanny, ChaConRW, ConMixing, MarPieEarly, MarPieTehCosmic}. 
In this sense, this principle is able to unify the Standard Model with general relativity under 
a generalized notion of Riemannian geometry.

Considering the variety of noncommutative geometric spaces and the physical 
implications of the spectral action \cite{ChaConUFNCG, ChaConConceptual, 
ChaConGravity, ChaConMarGS}, it is of great importance to use and develop 
different methods for computing this action. In particular, any interpretation of 
the full expansion of a spectral action is desirable. For example, using the Poisson 
summation formula, the spectral action for the Dirac operator on highly
symmetric manifolds, such as products of spheres by tori, was computed in 
\cite{ChaConUncanny}.

A considerable amount of work has been carried 
out on the cosmological implications of the spectral action 
in recent years  \cite{KolMar, Mar, MarPieEarly, 
MarPieTeh2012, MarPieTehCosmic, NelOchSal, NelSak1, 
NelSak2, EstMar}. For the Euclidean Robertson-Walker 
spacetime with a general cosmic factor $a(t)$, 
Chamseddine and Connes have devised an efficient 
method in \cite{ChaConRW} for computing the terms of 
the spectral action, which is 
based on making the use of the Euler-Maclaurin formula and 
the Feynman-Kac formula.  They computed the terms up 
to $a_{10}$  in the expansion and made a conjecture, which  
was addressed in \cite{FatGhoKha} by 
using pseudodifferential operators and heat kernel techniques.  
That is, it was shown  that a general term in the expansion 
is described by a polynomial with rational coefficients in $a(t)$ 
and its derivatives of a certain order.

The present paper is intended to study the 
spectral action for the Bianchi type-IX minisuperspace,  
which is a homogeneous \emph{anisotropic} 
cosmological model. Since the spectral action of a geometry 
depends on the eigenvalues of the square of its Dirac operator, 
we explicitly compute the Dirac operator $D$ of the Bianchi type-IX 
metric in \S \ref{DiracOperatorSection} and 
derive the pseudodifferential symbol of $D^2$. In \S 
\ref{TermsCalculationSection}, following a brief review of 
the heat kernel method that uses pseudodifferential  
calculus for the computation 
of the Seeley-de Witt coefficients \cite{GilBook1}, 
we present the calculation of the terms $a_0, a_2, a_4$ 
in the expansion of the spectral action associated with $D$.

We devise a new method for calculating the Seeley de-Witt 
coefficients of a geometry in terms of noncommutative residues 
of operators, which extends the result on the realization of the 
Einstein-Hilbert action as the residue of a power of the Laplacian, 
see \cite{KalWal, Kas, GraVarFig}.   
This method is explained in detail in \S \ref{WodzickiResidueSection} 
and it checks out to give the same result for the calculated terms 
$a_0, a_2, a_4$ for the Bianchi type-IX metric.  Combining the 
symmetries of the metric with technical properties of 
pseudodifferential symbols of parametrices of the Laplacians, 
which significantly simplify in view of the new method using the 
Wodzicki residue, we prove a rationality result for a general term in 
the expansion of the spectral action for the Bianchi type-IX metric in 
\S \ref{RationalitySection}. That is, we show that  general terms  
of the expansion are expressed by several variable polynomials with rational 
coefficients evaluated on $w_1(t), w_2(t), w_3(t),$ and their derivatives 
of certain orders. In \S \ref{ThetaSection}, we discuss the gravitational 
instantons, which form an especially interesting class of Bianchi type-IX models, 
and elaborate on the significance of the rationality for the spectral action in 
relation to the arithmetic and number theoretic structures in mathematical physics. 
Finally, our main results and conclusions 
are summarized in \S \ref{ConclusionsSection}.

\section{The Dirac Operator of Bianchi Type-IX Metrics} 
\label{DiracOperatorSection}

The heat kernel method that uses pseudodifferential calculus for computing the 
Seeley-de Witt coefficients of an elliptic operator on a compact 
manifold relies on the pseudodifferential symbol 
of the operator in local charts \cite{GilBook1}. Thus, in this section we 
compute the Dirac operator $D$ of the  Bianchi type-IX metric 
and thereby obtain the symbol of $D^2$.

The most efficient way of computing the Dirac operator of a 
geometric space is to use an orthonormal coframe $\{ \theta^a \}$ 
for the metric, which, from the definition of $D$, 
yields  
\[
D = \sum_a \theta^a \nabla^S_{\theta_a}, 
\]
where $\nabla^S$ is the spin connection of the spin bundle and $\{ \theta_a \}$ 
is the predual of the coframe, see \cite{FriBook}.  Since $\nabla^S$ is the lift of the Levi-Civita 
connection $\nabla$ to the spin bundle, one starts with computing the matrix of 1-forms 
$\omega = (\omega^a_b) $ such that $\nabla = d + \omega$  in terms of $\theta^a$ 
in a local chart $(x^\mu)$, which can be lifted to the 
matrix of  the spin connection 1-forms  by making use of the 
Lie algebra isomorphism  $\mu:\mathfrak{so}(m)\to \mathfrak{spin}(m)$ given by 
\begin{equation} \nonumber
\mu(A)= \frac{1}{4}\sum_{a,b} A^{a b} e_a e_b, 
\qquad A = (A^{ab})\in \mathfrak{so}(m), 
\end{equation}
where $m$ is the dimension of the manifold.  
We note that $\{ e_a \}$ is the standard basis for $\mathbb{R}^m$ 
considered inside the Clifford algebra of $\mathbb{R}^m$ where  $\mathfrak{spin}(m)$ 
is spanned linearly by $\{ e_a e_b; a<b \}$.

The $\omega^a_b$ are found uniquely by writing 
\[
\nabla \theta^a = 
\sum_b \omega^a_b \otimes \theta^b, 
\]
and imposing the conditions that characterize the Levi-Civita 
connection, namely metric-compatibility 
and torsion-freeness which respectively imply that 
\[
\omega^a_b = 
- \omega^b_a, \qquad d \theta^a = 
\sum_b \omega^a_b \wedge \theta^b. 
\] 
Therefore the Dirac operator is written as 
\[
D = \sum_{a, \mu} \gamma^a dx^{\mu}(\theta_a) \frac{\partial}{\partial x^{\mu}} + 
\frac{1}{4} \sum_{a, b, c} \gamma^c \omega^b_{ac} \gamma^a \gamma^b, 
\]
where $\omega^b_{ac}$ are defined by 
\[
\omega^b_a = \sum_c \omega^b_{ac} \theta^c, 
\]  
and the matrices $\gamma^a$ represent the Clifford action of $\theta^a$ on the 
spin bundle, namely that they satisfy the relations $(\gamma^a)^2 = - I$ 
and $\gamma^a \gamma^b + \gamma^b \gamma^a = 0$ for $a \neq b$.

We go through this process for the Bianchi type-IX metric, 
\begin{eqnarray}
ds^2 &=& w_1 w_2 w_3 \, dt \,dt+\frac{w_1 w_2 \cos (\eta )}{w_3}d\phi \,d\psi 
   +\frac{w_1 w_2 \cos (\eta )}{w_3} d\psi \,d\phi \nonumber \\ 
 &   +&\left(\frac{w_2 w_3 \sin ^2(\eta )
   \cos ^2(\psi )}{w_1}+w_1 \left(\frac{w_3 \sin ^2(\eta ) \sin ^2(\psi )}{w_2}+\frac{w_2
   \cos ^2(\eta )}{w_3}\right)\right) d\phi \,d\phi  \nonumber \\ 
&  +& \frac{\left(w_1^2-w_2^2\right) w_3
   \sin (\eta ) \sin (\psi ) \cos (\psi )}{w_1 w_2} d\eta \,d\phi \nonumber \\ 
& + &\frac{\left(w_1^2-w_2^2\right) w_3 \sin (\eta ) \sin (\psi ) \cos (\psi )}{w_1
   w_2}d\phi \,d\eta  \nonumber \\ 
 &  +&\left(\frac{w_2 w_3 \sin ^2(\psi )}{w_1}  +\frac{w_1 w_3 \cos
   ^2(\psi )}{w_2}\right)d\eta \,d\eta +\frac{w_1 w_2}{w_3}d\psi \,d\psi, \nonumber
\end{eqnarray}
which is written in the local coordinates $(x^{\mu}) = (t, \eta, \phi, \psi)$, where 
$\mathbb{S}^3$ is parametrized by the map 
\[
( \eta, \phi, \psi) 
\mapsto 
\left ( \cos(\eta/2) e^{i (\phi+\psi)/2},   \sin(\eta/2) e^{i (\phi-\psi)/2}  \right ), 
\] 
with the parameter ranges $0 \leq \eta \leq \pi, 0 \leq \phi < 2 \pi, 0 \leq \psi < 4 \pi$. 
An orthonormal coframe for $ds^2$ is given by 
\begin{eqnarray}
\theta^0 &= &\sqrt{w_1 w_2 w_3}\,dt, \nonumber \\ 
\theta^1 &=&  \sin (\eta ) \cos (\psi ) \sqrt{\frac{w_2
   w_3}{w_1}}\,d\phi -\sin (\psi )
   \sqrt{\frac{w_2 w_3}{w_1}} \,d\eta, 
\nonumber \\ 
\theta^2 &=& \sin (\eta ) \sin (\psi ) \sqrt{\frac{w_1
   w_3}{w_2}}\,d\phi +\cos (\psi )
   \sqrt{\frac{w_1 w_3}{w_2}}\,d\eta, \nonumber \\ 
\theta^3 &=& \cos (\eta ) \sqrt{\frac{w_1
   w_2}{w_3}}\,d\phi + \sqrt{\frac{w_1
   w_2}{w_3}} \,d\psi.  \nonumber
\end{eqnarray}

By explicit calculations in this basis we find that  
the non-vanishing $\omega^b_{ac}$ are determined 
by the following terms: 
\begin{align*}
\omega^0_{11} &= -\frac{w_2 \left( w_1 w_3'-w_3
   w_1' \right)+w_1 w_3 w_2'}{2
   (w_1 w_2w_3)^{3/2}}, 
&      
\omega^0_{22} &= 
-\frac{w_2 \left(w_3 w_1'+w_1
   w_3'\right)-w_1 w_3w_2'}
   {2(w_1w_2 w_3)^{3/2}}, \\
 \omega^0_{33} &= 
-\frac{w_2 \left( w_3 w_1'-w_1
   w_3'\right)+w_1w_3 w_2'}{2
   (w_1w_2w_3)^{3/2}}, &
\omega^1_{23}&=
-\frac{w_1^2 w_2^2-w_3^2
   \left(w_1^2+w_2^2\right)}
   {2(w_1w_2 w_3)^{3/2}}, \\
 \omega^1_{32} &=
-\frac{w_1^2
   \left(w_2^2-w_3^2\right)+w_2^2
   w_3^2}
   {2(w_1w_2 w_3)^{3/2}}, &
\omega^2_{31}&=
-\frac{w_2^2 w_3^2-w_1^2
   \left(w_2^2+w_3^2\right)}{2(w_1w_2 w_3)^{3/2}}.
\end{align*}
Thus we achieve an explicit calculation of the pseudodifferential symbol of the 
Dirac operator which is written as 
\begin{eqnarray*}
\sigma(D)(x, \xi) &=& 
\sum_{a, \mu} i\gamma^{a} e_{a}^{\mu}\xi_{\mu+1}
+
\frac{1}{4\sqrt{w_{1}w_{2}w_{3}}}\left(\frac{w_{1}^{'}}{w_{1}}
+\frac{w_{2}^{'}}{w_{2}}+\frac{w_{3}^{'}}{w_{3}}\right)\gamma^{1} \\
&&-\frac{\sqrt{w_{1}w_{2}w_{3}}}{4}\left(\frac{1}{w_{1}^{2}}
+\frac{1}{w_{2}^{2}}+\frac{1}{w_{3}^{2}}\right)
\gamma^{2}\gamma^{3} \gamma^4 \\
&=&-\frac{i  \gamma^2
   \sqrt{w_1} \left(\csc (\eta )
   \cos (\psi ) \left(\xi _4 \cos (\eta
   )-\xi _3\right)+\xi _2 \sin (\psi
   )\right)}{\sqrt{w_2}
   \sqrt{w_3}} \\ 
&&+\frac{i 
   \gamma^3 \sqrt{w_2} \left(\sin
   (\psi ) \left(\xi _3 \csc (\eta )-\xi _4
   \cot (\eta )\right) +\xi _2 \cos (\psi
   )\right)}{\sqrt{w_1}
   \sqrt{w_3}} \\ 
&&+\frac{i 
   \gamma^1 \xi _1}{\sqrt{w_1}
   \sqrt{w_2}
   \sqrt{w_3}}+\frac{i 
   \gamma^4 \xi _4
   \sqrt{w_3}}{\sqrt{w_1}
   \sqrt{w_2}} \\ 
&&+ 
\frac{1}{4\sqrt{w_{1}w_{2}w_{3}}}\left(\frac{w_{1}^{'}}{w_{1}}
+\frac{w_{2}^{'}}{w_{2}}+\frac{w_{3}^{'}}{w_{3}}\right)\gamma^{1} \\
&&-\frac{\sqrt{w_{1}w_{2}w_{3}}}{4}\left(\frac{1}{w_{1}^{2}}+\frac{1}{w_{2}^{2}}+\frac{1}{w_{3}^{2}}\right)
\gamma^{2}\gamma^{3} \gamma^4, \\ 
\end{eqnarray*}
where the following non-vanishing $e_{a}^{\mu}$ are  used: 
\begin{align*}
e^0_0&=\frac{1}{\sqrt{w_1w_2w_3}}, &
e^1_1&=-\frac{\sqrt{w_1} \sin (\psi )}{\sqrt{w_2 w_3}}, &
e^1_2&=\frac{\sqrt{w_2} \cos (\psi )}{\sqrt{w_1w_3}}, \\
e^2_1&= \frac{\sqrt{w_1} \csc (\eta ) \cos (\psi )}{\sqrt{w_2 w_3}}, &    
e^2_2 &= \frac{\sqrt{w_2} \csc (\eta ) \sin (\psi )}{\sqrt{w_1w_3}}, &
e^3_1&=-\frac{\sqrt{w_1} \cot (\eta ) \cos (\psi )}{\sqrt{w_2 w_3}}, \\
e^3_2&=-\frac{\sqrt{w_2} \cot (\eta ) \sin (\psi )}{\sqrt{w_1w_3}}, & 
e^3_3&=\frac{\sqrt{w_3}}{\sqrt{w_1w_2}}. 
\end{align*}

Having the symbol of $D$, a direct calculation yields 
\[
\sigma(D^2)(x, \xi)= p_2(x, \xi) + p_1(x, \xi) + p_0(x, \xi), 
\]
where the expressions for the $p_k(x, \xi)$ are recorded below. 
The principal symbol of $D^2$, which is homogeneous of order 
2 in $\xi$, is given by 
\begin{eqnarray*}
p_2(x, \xi)&= &  
\frac{1}{w_1 w_2 w_3} \Large (
\xi _4^2 w_1^2 \cot ^2(\eta ) \cos
   ^2(\psi )+\xi _3^2 w_1^2 \csc
   ^2(\eta ) \cos ^2(\psi ) 
\\ &&
+\xi _2 \xi _4
   w_1^2 \cot (\eta ) \sin (2 \psi
   )-\xi _2 \xi _3 w_1^2 \csc (\eta
   ) \sin (2 \psi ) 
 \\ 
 &&
 -2 \xi _3 \xi _4
   w_1^2 \cot (\eta ) \csc (\eta )
   \cos ^2(\psi )+\xi _2^2 w_1^2
   \sin ^2(\psi )
\\&&
   +\xi _4^2 w_2^2
   \cot ^2(\eta ) \sin ^2(\psi )-\xi _2 \xi
   _4 w_2^2 \cot (\eta ) \sin (2
   \psi )
\\&&
+\xi _3^2 w_2^2 \csc
   ^2(\eta ) \sin ^2(\psi )+\xi _2 \xi _3
   w_2^2 \csc (\eta ) \sin (2 \psi
   ) \\&&-2 \xi _3 \xi _4 w_2^2 \cot
   (\eta ) \csc (\eta ) \sin ^2(\psi )+\xi
   _2^2 w_2^2 \cos ^2(\psi )
\\&&
+\xi_4^2 w_3^2+\xi _1^2 \Large ) I, 
\end{eqnarray*}
where  $I$ is the $4 \times 4$ identity matrix. 
The component of  $\sigma(D^2)$ that is homogeneous 
of order 1 has a lengthy expression: 
\begin{eqnarray*}
p_1(x, \xi) &=& 
\Large ( -\frac{i \xi _2 w_1 \cot (\eta )
   \cos ^2(\psi )}{w_2
   w_3}-\frac{i \xi _2 w_2
   \cot (\eta ) \sin ^2(\psi )}{w_1
   w_3} 
   \\&&
   -\frac{3 i \xi _4
   w_2 \csc ^2(\eta ) \sin (2 \psi
   )}{4 w_1 w_3} +\frac{3 i
   \xi _4 w_1 \csc ^2(\eta ) \sin (2
   \psi )}{4 w_2
   w_3} 
   \\ && 
   -\frac{i \xi _4 w_2
   \cos (2 \eta ) \csc ^2(\eta ) \sin (2 \psi
   )}{4 w_1 w_3}+\frac{i
   \xi _4 w_1 \cos (2 \eta ) \csc
   ^2(\eta ) \sin (2 \psi )}{4 w_2
   w_3}
   \\&&
   +\frac{i \xi _3 w_2
   \cot (\eta ) \csc (\eta ) \sin (2 \psi
   )}{w_1 w_3}-\frac{i \xi
   _3 w_1 \cot (\eta ) \csc (\eta )
   \sin (2 \psi )}{w_2 w_3} \Large ) I 
   \\&& 
   + \Large ( \frac{i \xi _4 w_3}{2
   w_1^2}+\frac{i \xi _4
   w_3}{2 w_2^2}-\frac{i
   \xi _4}{2 w_3} \Large ) \gamma^2 \gamma^3 
   \\ && 
   + \Large ( \frac{i \xi _4 w_2 \cot (\eta )
   \sin (\psi )}{2 w_1^2}-\frac{i
   \xi _3 w_2 \csc (\eta ) \sin
   (\psi )}{2 w_1^2}-\frac{i \xi _2
   w_2 \cos (\psi )}{2
   w_1^2} 
   \\ &&
   +\frac{i \xi _4
   w_2 \cot (\eta ) \sin (\psi )}{2
   w_3^2}-\frac{i \xi _3
   w_2 \csc (\eta ) \sin (\psi )}{2
   w_3^2}-\frac{i \xi _2
   w_2\cos (\psi )}{2
   w_3^2} 
   \\&&
   -\frac{i \xi _4 \cot (\eta
   ) \sin (\psi )}{2 w_2}+\frac{i
   \xi _3 \csc (\eta ) \sin (\psi )}{2
   w_2}+\frac{i \xi _2 \cos (\psi
   )}{2 w_2}
   \Large ) \gamma^2 \gamma^4 
   \\&& 
  + \Large (-\frac{i \xi _4 w_1 \cot (\eta )
   \cos (\psi )}{2 w_2^2}+\frac{i
   \xi _3 w_1 \csc (\eta ) \cos
   (\psi )}{2 w_2^2}-\frac{i \xi _2
   w_1 \sin (\psi )}{2
   w_2^2} 
   \\ &&
   -\frac{i \xi _4
   w_1 \cot (\eta ) \cos (\psi )}{2
   w_3^2}+\frac{i \xi _3
   w_1 \csc (\eta ) \cos (\psi )}{2
   w_3^2}-\frac{i \xi _2
   w_1 \sin (\psi )}{2
   w_3^2} 
   \\ &&
   +\frac{i \xi _4 \cot (\eta
   ) \cos (\psi )}{2 w_1}-\frac{i
   \xi _3 \csc (\eta ) \cos (\psi )}{2
   w_1}+\frac{i \xi _2 \sin (\psi
   )}{2 w_1} \Large ) \gamma^3 \gamma^4 
   \\&&
   + \Large ( -\frac{i \xi _4 w_1'}{2
   w_1^2 w_2}-\frac{i \xi
   _4 w_2'}{2 w_1
   w_2^2}+\frac{i \xi _4
   w_3'}{2 w_1
   w_2 w_3}  \Large ) \gamma^1 \gamma^4 
   \\&&
  + \Large ( -\frac{i \xi _4 \cot (\eta ) \cos (\psi )
   w_1'}{2 w_1
   w_2 w_3}+\frac{i \xi _3
   \csc (\eta ) \cos (\psi )
   w_1'}{2 w_1
   w_2 w_3}-\frac{i \xi _2
   \sin (\psi ) w_1'}{2
   w_1 w_2
   w_3} 
   \\ &&
   +\frac{i \xi _4 \cot (\eta )
   \cos (\psi ) w_2'}{2
   w_2^2 w_3}-\frac{i \xi
   _3 \csc (\eta ) \cos (\psi )
   w_2'}{2 w_2^2
   w_3}+\frac{i \xi _2 \sin (\psi )
   w_2'}{2 w_2^2
   w_3} 
   \\&&
   +\frac{i \xi _4 \cot (\eta )
   \cos (\psi ) w_3'}{2
   w_2 w_3^2}-\frac{i \xi
   _3 \csc (\eta ) \cos (\psi )
   w_3'}{2 w_2
   w_3^2}+\frac{i \xi _2 \sin (\psi
   ) w_3'}{2 w_2
   w_3^2} \Large ) \gamma^1 \gamma^2 
   \\ && 
  + \Large ( 
   \frac{i \xi _4 \cot (\eta ) \sin (\psi )
   w_1'}{2 w_1^2
   w_3}-\frac{i \xi _3 \csc (\eta )
   \sin (\psi ) w_1'}{2
   w_1^2 w_3}-\frac{i \xi
   _2 \cos (\psi ) w_1'}{2
   w_1^2 w_3} 
   \\&& 
   -\frac{i \xi
   _4 \cot (\eta ) \sin (\psi )
   w_2'}{2 w_1
   w_2 w_3}+\frac{i \xi _3
   \csc (\eta ) \sin (\psi )
   w_2'}{2 w_1
   w_2 w_3}+\frac{i \xi _2
   \cos (\psi ) w_2'}{2
   w_1 w_2
   w_3} 
   \\&&
   +\frac{i \xi _4 \cot (\eta )
   \sin (\psi ) w_3'}{2
   w_1 w_3^2}-\frac{i \xi
   _3 \csc (\eta ) \sin (\psi )
   w_3'}{2 w_1
   w_3^2}-\frac{i \xi _2 \cos (\psi
   ) w_3'}{2 w_1
   w_3^2}
   \Large ) \gamma^1 \gamma^3. 
\end{eqnarray*}
Finally we have the zero order part of $\sigma(D^2)$: 
\begin{eqnarray*}
p_0(x, \xi) &=& 
\Large( -\frac{w_1'}{8 w_1
   w_2^2}-\frac{w_1'}{8
   w_1 w_3^2}+\frac{3
   w_1'}{8
   w_1^3}-\frac{w_2'}{8
   w_1^2
   w_2}-\frac{w_3'}{8
   w_1^2
   w_3}-\frac{w_2'}{8
   w_2 w_3^2} 
   \\&&
   +\frac{3
   w_2'}{8
   w_2^3}-\frac{w_3'}{8
   w_2^2 w_3}+\frac{3
   w_3'}{8 w_3^3} \Large) \gamma^1 \gamma^2 \gamma^3 \gamma^4 + 
   \\ && 
   \Large ( -\frac{w_1''}{4 w_1^2
   w_2
   w_3}+\frac{w_1'
   w_2'}{8 w_1^2
   w_2^2
   w_3}+\frac{w_1'
   w_3'}{8 w_1^2
   w_2 w_3^2}+\frac{5
   w_1'^2}{16 w_1^3
   w_2
   w_3}-\frac{w_2''}{4
   w_1 w_2^2
   w_3} 
   \\ &&
   +\frac{w_2'
   w_3'}{8 w_1
   w_2^2 w_3^2}+\frac{5
   w_2'^2}{16 w_1
   w_2^3
   w_3}-\frac{w_3''}{4
   w_1 w_2
   w_3^2}+\frac{5
   w_3'^2}{16 w_1
   w_2
   w_3^3}+\frac{w_2
   w_3}{16
   w_1^3} 
   \\&&
   +\frac{w_3}{8
   w_1
   w_2}+\frac{w_1
   w_3}{16
   w_2^3}+\frac{w_2}{8
   w_1
   w_3}+\frac{w_1}{8
   w_2
   w_3}+\frac{w_1
   w_2}{16 w_3^3} \Large ) I. 
\end{eqnarray*}

\section{Calculation of the terms up to $a_4$ in the Spectral Action} 
\label{TermsCalculationSection}

Calculation of the Seeley-de Witt coefficients associated with 
an elliptic positive differential operator on an $m$-dimensional 
compact manifold $M$ can be achieved by using the Cauchy 
integral formula and parametric pseudodifferential calculus,  
see \cite{GilBook1} for a detailed discussion. Let us review this 
method briefly for the operator $D^2$, where $D$ is the Dirac 
operator acting on a spin bundle on $M$.   In order to derive 
a small time asymptotic expansion of the form 
\[
\text{Trace}\big( e^{-t D^2}   \big ) 
\sim 
t^{m/2} \sum_{n=0}^\infty a_{2n}(D^2) t^n \qquad ({t \to 0^+}), 
\] 
one can start with the Cauchy integral formula to write 
\[
e^{-t D^2} = \frac{1}{2 \pi i} \int_\gamma e^{-t \lambda} (D^2 - \lambda )^{-1} \, d\lambda, 
\]
where the contour  $\gamma$ in the complex plane goes around the non-negative 
real numbers clockwise. Then, the idea is  to approximate 
$(D^2 - \lambda )^{-1}$ by pseudodifferential operators and to derive 
the above expansion by computing the trace of the corresponding 
approximation of the heat kernel.

The symbol of $D^2$ is of the form  
$p_2(x, \xi)+ p_1(x, \xi) + p_0(x, \xi)$ where each $p_i$ 
is homogeneous of order $i$ in $\xi$.  Since $D^2$ is an elliptic differential 
operator of order 2, the inverse of $D^2 - \lambda$ is approximated 
by its parametrix $R_\lambda$ with 
\[
\sigma(R_\lambda) \sim \sum_{j=0}^\infty r_j (x, \xi, \lambda), 
\]
where each $r_j (x, \xi, \lambda)$ is a parametric pseudodifferential 
symbol of order $-2 - j$, in the sense that 
\[
r_j(x, t \xi, t^2 \lambda) = t^{-2-j} r_j(x,  \xi,  \lambda). 
\]
The equation 
\[
\sigma((D^2 - \lambda) R_\lambda ) 
\sim
( \left (p_2(x, \xi)-\lambda) + 
p_1(x, \xi) + p_0(x, \xi) \right ) \circ 
\left ( \sum_{j=0}^\infty r_j(x, \xi, \lambda) \right) \sim 1
\]
can be solved recursively by comparing the homogeneous 
terms on the both sides after expanding the left hand side 
using the asymptotic composition rule for the symbols. Indeed, 
one finds that 
\[
r_0(x, \xi, \lambda) = (p_2(x, \xi) - \lambda)^{-1}, 
\]
and for any $n >1$, the term $r_n(x, \xi, \lambda)$ is found to 
be expressed in terms of  
$r_0(x, \xi, \lambda), \dots, r_{n-1}(x, \xi, \lambda) $ by the formula 
\[
r_n(x, \xi, \lambda) = - \sum \frac{1}{\alpha !} 
\partial_\xi^\alpha r_j(x, \xi, \lambda) \, D_x^\alpha p_k(x, \xi) \, r_0(x, \xi, \lambda),  
\]
where the summation is over all 
$\alpha \in \mathbb{Z}_{\geq 0}^4, j \in \{ 0, 1, \dots , n-1\}, k \in \{0, 1, 2 \},$ 
such that $|\alpha|+j +2 -k =n$.

The coefficients $a_{2n}(D^2)$ in the small time asymptotic expansion 
are then computable by 
integrating the invariantly defined functions 
\[
a_{2n}(x, D^2)= 
\frac{(2 \pi)^{-m}}{2 \pi i} \int \int_{\gamma} e^{-\lambda} \, \textnormal{tr} 
\left ( r_{2n}(x, \xi, \lambda) \right )  \, d \lambda \, d^m \xi 
\]
over the manifold against the volume form, which shows that these coefficients are local 
invariants of the geometry. We note that the odd coefficients vanish 
since for any odd $j$, the term $r_j(x, \xi, \lambda)$ is an odd function 
of the variable $\xi \in \mathbb{R}^m$, whose integral over $\mathbb{R}^m$ 
vanishes.

Applying this method to Bianchi type-IX metric we compute 
the corresponding $a_0, a_2, a_4$, which are recorded below 
(without writing the integral with respect $t$).  The volume term 
is simply 
\begin{eqnarray*}
a_0(D^2) &=&4w_{1}w_{2}w_{3}, 
\end{eqnarray*}
and the scalar curvature term, after remarkable cancellations, is found to be  
\begin{eqnarray*}
a_2(D^2) &=&
-\frac{w_1^2}{3}-\frac{w_2^2}{3}
-\frac{w_3^2}{3}+\frac{w_1^2 w_2^2}{6 w_3^2}
+\frac{w_1^2 w_3^2}{6 w_2^2}+\frac{w_2^2 w_3^2}{6 w_1^2}
-\frac{\left(w_1'\right){}^2}{6 w_1^2}-\frac{\left(w_2'\right){}^2}{6 w_2^2} 
\\ &&-\frac{\left(w_3'\right){}^2}{6 w_3^2}-\frac{w_1' w_2'}{3 w_1 w_2}-\frac{w_1' w_3'}{3  w_1
   w_3}-\frac{w_2' w_3'}{3 w_2 w_3}+\frac{w_1''}{3 w_1}+\frac{w_2''}{3 w_2}+\frac{w_3''}{3 w_3}. 
\end{eqnarray*}
Although it seems lengthy, after an enormous amount of 
cancellations, the next coefficient is expressed as:

\begin{eqnarray*}
a_4(D^2)&=& 
   -\frac{w_1^3 w_2^3}{15 w_3^5}-\frac{w_1^3 w_3^3}{15 w_2^5}
   -\frac{w_2^3 w_3^3}{15 w_1^5}+\frac{w_1^3 w_2}{15
   w_3^3}+\frac{w_1 w_2^3}{15 w_3^3}
   +\frac{w_1^3 w_3}{15 w_2^3}+\frac{w_2^3 w_3}{15 w_1^3}
   +\frac{w_1 w_3^3}{15
   w_2^3} 
\end{eqnarray*}   
\begin{eqnarray*}
&&
   +\frac{w_2 w_3^3}{15 w_1^3}
   -\frac{w_1 w_2}{15 w_3}-\frac{w_1 w_3}{15 w_2}
   -\frac{w_2 w_3}{15w_1}-\frac{w_2 \left(w_1'\right){}^2}{15 w_1 w_3^3}
   -\frac{w_3 \left(w_1'\right){}^2}{15 w_1 w_2^3}-\frac{w_3
   \left(w_2'\right){}^2}{15 w_1^3 w_2} 
   \\&&
   -\frac{w_1 \left(w_2'\right){}^2}{15 w_2 w_3^3}
   -\frac{w_1
   \left(w_3'\right){}^2}{15 w_2^3 w_3}-\frac{w_2 \left(w_3'\right){}^2}{15 w_1^3
   w_3}+\frac{2\left(w_1'\right){}^2}{15 w_1 w_2 w_3}+\frac{2\left(w_2'\right){}^2}{15 w_1 w_2
   w_3}
   \\&&
   +\frac{2\left(w_3'\right){}^2}{15 w_1 w_2 w_3}
   -\frac{w_2 \left(w_1'\right){}^2}{18 w_1^3 w_3} 
   -\frac{w_3
   \left(w_1'\right){}^2}{18 w_1^3 w_2}-\frac{w_1 \left(w_2'\right){}^2}{18 w_2^3 w_3} 
   -\frac{w_3
   \left(w_2'\right){}^2}{18 w_1 w_2^3} 
   \\&&
   -\frac{w_1 \left(w_3'\right){}^2}{18 w_2 w_3^3}-\frac{w_2
   \left(w_3'\right){}^2}{18 w_1 w_3^3}-\frac{w_2 w_3 \left(w_1'\right){}^2}{18 w_1^5}-\frac{w_1 w_3
   \left(w_2'\right){}^2}{18 w_2^5}-\frac{w_1 w_2 \left(w_3'\right){}^2}{18 w_3^5} 
   \\&&
   -\frac{31
   \left(w_1'\right){}^4}{90 w_1^5 w_2 w_3}-\frac{31 \left(w_2'\right){}^4}{90 w_1 w_2^5 w_3}-\frac{31
   \left(w_3'\right){}^4}{90 w_1 w_2 w_3^5}-\frac{7 w_1' w_2'}{60 w_3^3} 
   -\frac{7 w_1' w_3'}{60 w_2^3}-\frac{7
   w_2' w_3'}{60 w_1^3} 
   \end{eqnarray*}
   \begin{eqnarray*} 
   &&
   -\frac{w_1' w_2'}{45 w_1^2 w_3}
   -\frac{w_1' w_2'}{45 w_2^2 w_3}
   -\frac{w_2' w_3'}{45 w_1
   w_3^2}
   +\frac{5 w_3 w_1' w_2'}{36 w_1^4}
   +\frac{5 w_3 w_1' w_2'}{36 w_2^4}
   +\frac{5 w_2 w_1' w_3'}{36
   w_1^4}
   \\&&
   +\frac{5 w_2 w_1' w_3'}{36 w_3^4}
   +\frac{5 w_1 w_2' w_3'}{36 w_2^4}
   +\frac{5 w_1 w_2' w_3'}{36
   w_3^4}
   +\frac{7 w_3 w_1' w_2'}{90 w_1^2 w_2^2}
   +\frac{7 w_2 w_1' w_3'}{90 w_1^2 w_3^2}
   +\frac{7 w_1 w_2'w_3'}{90 w_2^2 w_3^2}
   \\&&
   -\frac{41 \left(w_1'\right){}^3 w_2'}{180 w_1^4 w_2^2 w_3} 
   -\frac{41 w_1' \left(w_2'\right){}^3}{180 w_1^2 w_2^4 w_3}
   -\frac{41 \left(w_1'\right){}^3 w_3'}{180 w_1^4 w_2 w_3^2}-
   \frac{41
   w_1' \left(w_3'\right){}^3}{180 w_1^2 w_2 w_3^4}-
   \frac{41 w_2' \left(w_3'\right){}^3}{180 w_1 w_2^2
   w_3^4} 
   \\&&
   -\frac{41 \left(w_2'\right){}^3 w_3'}{180 w_1 w_2^4 w_3^2}
   -\frac{23 \left(w_1'\right){}^2
   \left(w_2'\right){}^2}{90 w_1^3 w_2^3 w_3}
   -\frac{23 \left(w_1'\right){}^2 \left(w_3'\right){}^2}{90 w_1^3 w_2
   w_3^3}
   -\frac{23 \left(w_2'\right){}^2 \left(w_3'\right){}^2}{90 w_1 w_2^3 w_3^3}
   \\&&
   -\frac{w_1' w_3'}{45 w_1^2
   w_2}-
   \frac{w_1' w_3'}{45 w_2 w_3^2} 
   -\frac{w_2' w_3'}{45 w_1 w_2^2}-\frac{91 \left(w_1'\right){}^2 w_2'
   w_3'}{180 w_1^3 w_2^2 w_3^2}-\frac{91 w_1' \left(w_2'\right){}^2 w_3'}{180 w_1^2 w_2^3 w_3^2}
   \\&&
   -\frac{91 w_1'
   w_2' \left(w_3'\right){}^2}{180 w_1^2 w_2^2 w_3^3}+\frac{w_2 w_1''}{24 w_3^3}+\frac{w_3 w_1''}{24
   w_2^3}+\frac{w_1 w_2''}{24 w_3^3}+\frac{w_3 w_2''}{24 w_1^3}+\frac{w_1 w_3''}{24 w_2^3}+\frac{w_2 w_3''}{24 w_1^3}
   \\&&
   -\frac{w_1''}{12 w_2 w_3} -\frac{w_2''}{12 w_1
   w_3}-\frac{w_3''}{12 w_1 w_2}+\frac{w_2 w_1''}{36 w_1^2 w_3}+\frac{w_3 w_1''}{36 w_1^2 w_2}+\frac{w_1 w_2''}{36
   w_2^2 w_3} 
   \\&&
   -\frac{5 w_2 w_3 w_1''}{72 w_1^4}-\frac{5 w_1 w_3 w_2''}{72 w_2^4}-\frac{5 w_1 w_2 w_3''}{72
   w_3^4}+\frac{5 \left(w_1'\right){}^2 w_1''}{8 w_1^4 w_2 w_3}+\frac{5 \left(w_2'\right){}^2 w_2''}{8 w_1 w_2^4
   w_3} 
   \\&&
   +\frac{5 \left(w_3'\right){}^2 w_3''}{8 w_1 w_2 w_3^4} +\frac{71 w_1' w_2' w_1''}{180 w_1^3 w_2^2
   w_3}+\frac{71 w_1' w_2' w_2''}{180 w_1^2 w_2^3 w_3}+\frac{71 w_1' w_3' w_1''}{180 w_1^3 w_2 w_3^2}+\frac{71
   w_1' w_3' w_3''}{180 w_1^2 w_2 w_3^3} \\&&
   +\frac{71 w_2' w_3' w_3''}{180 w_1 w_2^2 w_3^3}+\frac{71 w_2' w_3'
   w_2''}{180 w_1 w_2^3 w_3^2}+\frac{41 \left(w_2'\right){}^2 w_1''}{360 w_1^2 w_2^3 w_3}+\frac{41
   \left(w_3'\right){}^2 w_1''}{360 w_1^2 w_2 w_3^3}+\frac{41 \left(w_2'\right){}^2 w_3''}{360 w_1 w_2^3
   w_3^2} 
   \\&&
   +\frac{41 \left(w_3'\right){}^2 w_2''}{360 w_1 w_2^2 w_3^3}+\frac{41 \left(w_1'\right){}^2 w_2''}{360
   w_1^3 w_2^2 w_3}+\frac{41 \left(w_1'\right){}^2 w_3''}{360 w_1^3 w_2 w_3^2}+\frac{11 w_2' w_3' w_1''}{36 w_1^2
   w_2^2 w_3^2}+\frac{11 w_1' w_3' w_2''}{36 w_1^2 w_2^2 w_3^2} \\&&
   +\frac{11 w_1' w_2' w_3''}{36 w_1^2 w_2^2
   w_3^2}-\frac{\left(w_1''\right){}^2}{6 w_1^3 w_2 w_3}-\frac{\left(w_2''\right){}^2}{6 w_1 w_2^3
   w_3}-\frac{\left(w_3''\right){}^2}{6 w_1 w_2 w_3^3}+\frac{w_3 w_2''}{36 w_1 w_2^2} 
   \\&&
   +\frac{w_1 w_3''}{36 w_2
   w_3^2}+\frac{w_2 w_3''}{36 w_1 w_3^2}-\frac{w_1'' w_2''}{15 w_1^2 w_2^2 w_3}-\frac{w_2'' w_3''}{15 w_1 w_2^2
   w_3^2}-\frac{w_1'' w_3''}{15 w_1^2 w_2 w_3^2} 
   -\frac{w_1'w_1{}^{(3)}}{6 w_1^3 w_2 w_3} 
    \\&&
   -\frac{w_2' w_2{}^{(3)}}{6 w_1 w_2^3 w_3}-\frac{w_3' w_3{}^{(3)}}{6 w_1 w_2
   w_3^3} -\frac{w_2' w_1{}^{(3)}}{10 w_1^2 w_2^2 w_3}-\frac{w_3' w_1{}^{(3)}}{10 w_1^2 w_2 w_3^2}-\frac{w_1'
   w_2{}^{(3)}}{10 w_1^2 w_2^2 w_3}
   -\frac{w_3' w_2{}^{(3)}}{10 w_1 w_2^2 w_3^2} 
   \\&&
   -\frac{w_1' w_3{}^{(3)}}{10 w_1^2
   w_2 w_3^2}-\frac{w_2' w_3{}^{(3)}}{10 w_1 w_2^2 w_3^2}+\frac{w_1{}^{(4)}}{30 w_1^2 w_2
   w_3}+\frac{w_2{}^{(4)}}{30 w_1 w_2^2 w_3}+\frac{w_3{}^{(4)}}{30 w_1 w_2 w_3^2}. 
\end{eqnarray*}

Following the conjecture of Chamseddine and Connes 
for Robertson-Walker metrics \cite{ChaConRW}, which was 
addressed in \cite{FatGhoKha}, 
the crucial observation to make at this stage is that all of 
the coefficients appearing in the above terms  are 
rational numbers. This indicates that the rationality 
result holds for the Bianchi type-IX metric, which is 
proved in \S \ref{RationalitySection}.

\section{Heat Coefficients and the Wodzicki Residue} 
\label{WodzickiResidueSection}

In this section we introduce a method for computing the Seeley-de Witt 
coefficients by making use of Wodzicki's noncommutative residue \cite{Wod, Wod2}. The 
advantage of this method is that it involves significantly less complexity 
in computations, thus, for instance, it illuminates the structure of the heat expansion 
of the Bianchi type-IX metric more elegantly.

Given a closed $m$-dimensional manifold $M$, the Wodzicki residue 
is the unique trace functional on the algebra of classical pseudodifferential 
operators acting on the smooth sections of a vector bundle over $M$ 
(up to multiplication by a constant).   The local symbol $\sigma$ 
of a classical pseudodifferential operator $P_\sigma$ of order $d \in \mathbb{Z}$ 
has an asymptotic expansion of the form 
\[
\sigma (x, \xi) 
\sim 
\sum_{j=0}^\infty \sigma_{d-j} (x, \xi) \qquad (\xi \to \infty),
\]
where each $\sigma_{d-j}$ is positively homogeneous of order 
$d-j$ in $\xi$. The noncommutative residue of the operator 
$P_\sigma$ is defined by 
\[
\textrm{Res}(P_\sigma) = 
\int_{S^*M} \textrm{tr} \left (\sigma_{-m}(x, \xi) \right ) \,  d^{m-1}\xi \, d^mx, 
\]
where $S^*M = \{ (x, \xi) \in T^*M; ||\xi||_g=1 \}$ is the 
cosphere bundle of $M$ and the integral is in fact the 
integral of the corresponding Wodzicki residue density over $M$, 
see \cite{Wod, Wod2, Kassel} for more details.

An alternative definition for Res, which is quite spectral, 
provides a link between the Seeley-de Witt 
coefficients and the noncommutative residue. That is, 
for any  pseudodifferential operator $P_\sigma$, 
the map that sends a complex number $s$ with a large 
enough real part to  $\textrm{Trace}(P_\sigma \Delta^{-s} ),$ 
where $\Delta$ is a Laplacian, 
has a meromorphic extension to the complex plane with 
at most simple poles at its singularities.  The noncommutative 
residue can be defined as the linear functional 
\[
P_\sigma 
\mapsto 
\textrm{res}_{ s=0 }\textrm{Trace}(P_\sigma \Delta^{-s} ), 
\]
which turns out to be a trace functional. Thus, 
there is a constant $c_m$ such that for any 
classical $P_\sigma$, we have 
\[
\textrm{Res}(P_\sigma) =  
\int_{S^*M} \sigma_{-m}(x, \xi) \,d^{m-1}\xi \, d^m x =  
c_m \left ( \textrm{res}_{ s=0 }\textrm{Trace}(P_\sigma \Delta^{-s} ) \right). 
\]
The constant $c_m$ can be computed easily as follows. 
The operator $\textrm{Res}(\Delta^{- m/2})$ 
is of order $-m$ and its principal symbol is 
given by 
\[
\sigma_P(\Delta^{- m/2}) = 
\sigma_P(\Delta^{-1})^{ m/2} = 
\sigma_P(\Delta)^{-m/2}, 
\]
which yields 
\[
\textrm{Res}(\Delta^{- m/2}) = 
\int_{S^*M} \textnormal{tr} \left ( \sigma_P(\Delta)^{-m/2} \right ) \, d^{m-1}\xi \, d^m x. 
\]
On the other hand, writing $\sigma_P(\Delta) = \left ( \sum_{i,j} g^{ij} \xi_i \xi_j \right ) I $, 
we have  
\begin{eqnarray}
 \textrm{res}_{ s=0 }\textrm{Trace}( \Delta^{-m/2} \Delta^{-s} ) &=& 
\textrm{res}_{ s=m/2 }\textrm{Trace}( \Delta^{-s} )  \nonumber 
\end{eqnarray}
\begin{eqnarray}
&=& 
\frac{1}{\Gamma(m/2)}\frac{(2 \pi)^{-m}}{2 \pi i} \int \int \int_{\gamma} e^{-\lambda} \,\textnormal{tr} 
\left ( (\sigma_P(\Delta) - \lambda  \right)^{-1} )  \, d \lambda \, d^m \xi \, d^m x \nonumber \\ 
&=&
\frac{(2 \pi)^{-m}}{\Gamma(m/2)} \,\textnormal{rk}(V) \int \int e^{-\sum_{i, j}g^{i j} \xi_i \xi_j}\, d^m \xi \, d^m x \nonumber \\ 
&=& \frac{(2 \pi)^{-m}}{\Gamma(m/2)} \, \textnormal{rk}(V) \int \sqrt{\frac{\pi^m}{\textrm{det}(g^{ij})}}  \, d^m x \nonumber \nonumber \\
&=& \frac{2^{-m} \pi^{-m/2}}{\Gamma(m/2)} \,\textrm{rk}(V) \int \textrm{det}^{-1/2}(g^{ij}) \, d^m x, \nonumber
\end{eqnarray}
where for the second identity, we have used the formula 
\eqref{crucialformula} proved below, for $n=0$. 
Therefore, we have 
\[
c_m =  
2^m \pi^{m/2}  \Gamma(m/2)
\frac{ \int \left ( \sum_{i,j} g^{ij} \xi_i \xi_j \right)^{-m/2} \, d^{m-1} \xi \, d^m x   }
{ \int \textrm{det}^{-1/2}(g^{ij}) \, d^m x }
= 2^{m+1} \pi^m. 
\]

Using the Mellin transform,  
$\lambda^{-s} = 
\frac{1}{\Gamma(s)} \int_0^\infty e^{-t \lambda} t^{s-1} \,dt, \lambda >0, $ 
one has 
\[
\textrm{Trace}(\Delta^{-s}) = 
\frac{1}{\Gamma(s)} \int_0^\infty 
\left ( \textrm{Trace}(e^{-t \Delta})- \textrm{dim ker}(\Delta) \right ) t^{s} \,\frac{dt}{t}. 
\] 
By breaking the interval of the integration in the latter to $[0, 1]$ and 
$(1, \infty)$, and by substituting  the small time asymptotic expansion,   
\[
\text{Trace}\big( e^{-t \Delta}   \big ) = 
t^{- m /2} \sum_{n=0}^N a_{2n} t^n + O(t^{- m/2+N+1}),  
\]
in the first part, one finds that 
\[
\textrm{res}_{s=m/2-n} \textrm{Trace}(\Delta^{-s}) = 
\frac{a_{2n}(\Delta)}{\Gamma(m/2-n)}, 
\]
for any non-negative integer $n \leq m/2 -1$, see \cite{GilBook1}. In particular we have 
\begin{equation} \label{crucialformula}
\textrm{res}_{s=1} \textrm{Trace}(\Delta^{-s}) = 
a_{m-2}(\Delta). 
\end{equation}
This observation yields the following assertion, which is 
used crucially in the sequel.

\begin{lemma} \label{cruciallemma}
If $\Delta$ is a Laplacian acting on the smooth sections of a vector bundle 
over an $m$-dimensional manifold, then 
\[
a_{m-2}(\Delta) 
=  
\frac{1}{c_m} \textnormal{Res}(\Delta^{-1}) 
= 
\frac{1}{2^{m+1} \pi^m} \textnormal{Res}(\Delta^{-1}). 
\]
\begin{proof}
It follows from the identity \eqref{crucialformula} and the fact that 
\begin{eqnarray*}
\textrm{res}_{s=1} \textrm{Trace}(\Delta^{-s}) 
= 
\textrm{res}_{s=0} \textrm{Trace}(\Delta^{-1}\Delta^{-s}) 
= 
\frac{1}{c_m} \textnormal{Res}(\Delta^{-1}) .
\end{eqnarray*} 
\end{proof}
\end{lemma}

Since we are mainly concerned with studying the 
spectral action for the Bianchi type-IX metric in this article, 
let us assume that $D$ is the Dirac operator on a 4-dimensional 
manifold. 
By applying Lemma \eqref{cruciallemma} to $ \Delta=D^2$, 
we have 
\begin{eqnarray*}
a_2(D^2)  =
\frac{1}{c_4} \textrm{Res}(D^{-2}) 
= \frac{1}{32 \pi^4} \int_{S^*M} \textnormal{tr} \left ( \sigma_{-4}(D^{-2}) \right ) \, d^3 \xi \, d^4x, 
\end{eqnarray*}
where  $\sigma_{-4}(D^{-2})$ is the homogeneous component 
of order $-4$ in the expansion of the symbol of the parametrix of 
$D^2$. In the following theorem, we show that  
the next coefficients $a_{2n}(D^2), n \geq 2$, can similarly be 
expressed as  noncommutative residues of Laplacians.

\begin{theorem} \label{computationtheorem}
Let $D$ be the Dirac operator on a 4-dimensional 
manifold. For any non-negative even integer $r$, 
we have 
\[
a_{2+r}(D^2) = \frac{1}{2^5\, \pi^{4+r/2}} \textnormal{Res}(\Delta^{-1}),
\]
where 
\[
\Delta = D^2 \otimes 1 + 1 \otimes \Delta_{\mathbb{T}^r},
\]
in which $\Delta_{\mathbb{T}^r}$ is the flat Laplacian on 
the $r$-dimensional torus $\mathbb{T}^r = \left ( \mathbb{R}/\mathbb{Z} \right )^r$. 

\begin{proof}

It follows from Lemma \ref{cruciallemma} that 
\begin{eqnarray*}
a_{2+r}(\Delta) =\frac{1}{c_{4+r}} \textnormal{Res} (\Delta^{-1}), 
\end{eqnarray*}
where $\sigma_{-4-r}(\Delta^{-1}) $ is the homogeneous term of order $-4-r$ 
in the expansion of the symbol of the parametrix of $\Delta$.

Since the metric on $\mathbb{T}^r$ is chosen to be flat, its volume 
term is evidently the only non-zero heat coefficient, which combined with  
the K\"unneth formula (cf. \cite{GilBook1}) implies that   
\begin{eqnarray*}
a_{2+r}((x, x'), \Delta) &=& a_{2+r}(x, D^2) a_0(x', \mathbb{T}^r) \\
&=& 2^{-r} \pi^{-r/2}a_{2+r}(x, D^2). 
\end{eqnarray*}
Therefore, 
\begin{eqnarray*}
a_{2+r}(D^2) &=& 
\frac{2^{r} \pi^{r/2}}{c_{4+r}} \textrm{Res}(\Delta^{-1})
 \\
&=& \frac{1}{2^5 \pi^{4+r/2}} \textrm{Res}(\Delta^{-1}). 
\end{eqnarray*}

\end{proof}

\end{theorem}

A direct consequence of this theorem provides an efficient 
method for computing the Seeley-de Witt coefficients 
with significantly less complexities in the calculations. It also  
yields an elegant proof of the rationality result for the Bianchi 
type-IX metric, which is presented in the following section.

\begin{corollary} \label{crucialcorollary}
Assuming the conditions and notations of Theorem \ref{computationtheorem} we have 
\[
a_{2+r}(D^2) 
= 
\frac{1}{2^5 \pi^{4+r/2}} \int \textnormal{tr} \left ( \sigma_{-4-r}(\Delta^{-1}) \right ) \, d^{3+r} \xi \, d^{4}x. 
\] 
\begin{proof}
It follows from the fact that if $\sigma(D^2) = p_2(x, \xi) + p_1(x, \xi) + p_0(x, \xi)$, where 
each $p_i$ is homogeneous of order $i$ in $\xi,$ then $\sigma(\Delta) = 
p'_2(x, \xi) + p_1(x, \xi) + p_0(x, \xi)$, with $p'_2(x, \xi) = p_2(x, \xi) + (\xi_5^2 + 
\cdots +\xi_{4+r}^2)I.$ Therefore the homogeneous terms $\sigma_{-2-j}(\Delta^{-1})$ 
of order $-2-j$ in the 
expansion $\sigma(\Delta^{-1}) \sim \sum_{j=0}^\infty  \sigma_{-2-j}(\Delta^{-1})$   
are independent of the coordinates of $\mathbb{T}^r$. 

\end{proof}
\end{corollary}

We confirm the validity of the coefficients $a_0, a_2, a_4,$ calculated  
for the Bianchi type-IX metric in \S \ref{TermsCalculationSection}  
by noting that the method devised in the present section produces the same 
expressions. We stress that in practice the new method is significantly more 
convenient since the expression that leads to a Seeley-de Witt 
coefficient simplifies when one considers its restriction to the corresponding 
cosphere bundle in order to compute the noncommutative residue.

The noncommutative residue was originally discovered in the 1-dimensional 
case by Adler \cite{Adl} and Manin \cite{Man}. Its coincidence with the Dixmier 
trace \cite{ConAction} on pseudodifferential operators of order $-m$ on an 
$m$-dimensional closed manifold indicates its applicability for explicit and 
convenient computations. It is 
also worth mentioning that a noncommutative residue developed 
for noncommutative tori \cite{FatWon, FatKha} simplified 
a purely noncommutative heat kernel computation significantly 
and clarified in \cite{Fat} the reason for mysterious and remarkable 
cancellations that occur in this type of computations.

\section{Rationality of the Spectral Action for Bianchi Type-IX Metrics} 
\label{RationalitySection}

The Seeley-de Witt coefficients $a_{2n}$ appearing in the 
expansion of the spectral action for the Bianchi type-IX metric 
are expressed in terms of several variable polynomials 
with \emph{rational} coefficients evaluated on the 
cosmic evolution factors $w_1(t), w_2(t), w_3(t),$ and 
their derivatives of certain orders. This extends the 
statement conjectured in \cite{ChaConRW} and addressed in \cite{FatGhoKha} 
for Robertson-Walker metrics to a homogeneous \emph{anisotropic} 
cosmological model.

In order to prove the rationality result for the Bianchi type-IX 
metric, similar to the treatment in \cite{FatGhoKha}, let us start with 
the crucial observation that the local 
forms $a_{2n}(x,D^2)d^3x$, where $D$ is the Dirac operator 
of this geometry, are invariant over the spatial manifold $\mathbb{S}^3$. 
This can be seen from  
the defining formula \eqref{Bianchimetric} for the metric, in which the left 
invariance of the 1-forms $\sigma_1, \sigma_2, \sigma_3,$ implies 
that  the metric is invariant under any diffeomorphism arising 
from left multiplication by an element of $SU(2)$.  Since the action 
is transitive and left multiplication by an element of  
$SU(2)$ is an isometry, any 
isometry-invariant function on $\mathbb{S}^3$ 
is independent of the spatial coordinates. In particular, 
the restriction of the kernel of  $e^{-t D^2}$ to the diagonal 
and consequently the differential forms 
$a_{2n}(x,D^2) \, d^4x$ are invariant, and  
if we set $a_{2n}(x,D^2)\, d^3x = \tilde{a}_{2n}(x, D^{2}) \, dvol_g$,  
where   $dvol_g$ is the volume form, then $\tilde{a}_{2n}(x, D^{2})$ is 
independent of the spatial coordinates for any $n$.

Furthermore, we can easily determine the general form of 
$\tilde a_{2n}(x, D^2)$ by applying the method devised 
in \S \ref{WodzickiResidueSection}, which is based on making 
use of the noncommutative residue, combined with the K\"unneth 
formula and restricting the computations to the cosphere bundle.   
In fact, writing $\sigma(\Delta^{-1}) \sim \sum_{j=-2}^{-\infty} \sigma_j(x, \xi)$, 
where each $\sigma_j$ is homogeneous of order $j$, one finds recursively that 
\begin{eqnarray*}
\sigma_{-2}(x, \xi) &=& p_2(x, \xi)^{-1},  \\
\sigma_{-2-n}(x, \xi) &=& - \sum \frac{1}{\alpha !} 
\partial_\xi^\alpha \sigma_j(x, \xi) \, D_x^\alpha p_k(x, \xi) \, \sigma_{-2}(x, \xi)  \qquad (n > 0), 
\end{eqnarray*}
where the summation is over all multi-indices 
of non-negative integers $\alpha$, $-2-n <  j \leq -2, 0 \leq k \leq 2, $ 
such that $|\alpha|-j-k=n$.

Thus, if we define $\zeta_{\mu+1} = \sum_{\nu} e_{\mu}^{\nu}\xi_{\nu+1}$, then it can  be shown by induction that   
\begin{equation} \label{structureformula}
\left.\sigma_{-2-n}(x, \xi)\right|_{S^*(M\times\mathbb{T}^{n-2})} = \left.\sigma_{-2-n}(x, \xi(\zeta))\right|_{\zeta\in \mathbb{S}^{n+1}} = (w_1w_2w_3)^{-\frac{3}{2}n}P_{n}(\zeta),
\end{equation}
for any integer $n\ge2$, where $P_{n}(\zeta)$ is a polynomial in $\zeta_1, \dots ,\zeta_{n+2}$, 
with the coefficients being matrices whose entries are in the algebra 
generated by rational numbers, trigonometric functions of the spatial 
coordinates, and $W_{i}^{(p)}$ where $i \in \{1,2,3\}$, $p \in \{0,1,\dots,n\}$.
This fact leads to the following statement about the 
general form of the coefficients $a_{2n}(D^2)$.

\begin{theorem}
For any non-negative integer $n$, the coefficient $a_{2n}(D^2)$ in 
the expansion of the spectral action 
for the Bianchi-type IX metric is of the form
\[
a_{2n} (D^2) 
= 
(w_1w_2w_3)^{1-3n}Q_{2n}\left(w_1, w_2, w_3, w_1', w_2', w_3', \dots, w_1^{(2n)},  w_2^{(2n)},  w_3^{(2n)} \right ), 
\]
where $Q_{2n}$ is a polynomial with rational coefficients.
\begin{proof}
It follows from Corollary \ref{crucialcorollary} that
\begin{eqnarray*}
a_{2+r}(x,D^2) 
&=& 
\frac{1}{2^{5}\pi^{4+r/2}}\int_{S^*(M\times\mathbb{T}^r)}\textnormal{tr}(\sigma_{-4-r}(\Delta^{-1})) \, d^{3+r}\xi
\\
&=&
\frac{1}{2^{5}\pi^{4+r/2}}\int_{\mathbb{S}^{3+r}}\textnormal{tr}(\sigma_{-4-r}(\Delta^{-1})) \,dvol_g \, d^{3+r}\zeta,
\end{eqnarray*}
where $\zeta_{\mu+1} = \sum_{\nu} e_{\mu}^{\nu}\xi_{\nu+1}$ so 
that the Jacobian of the coordinate transformation is just 
$dvol_g$. This implies that 
\[
\tilde{a}_{2n}(x, D^{2}) 
= 
\frac{1}{2^{5}\pi^{n+3}}\int_{\mathbb{S}^{2n+1}}\textnormal{tr}(\sigma_{-2-2n}(\Delta^{-1})) \, d^{2n+1}\zeta,
\] 
which, as shown above, is independent of the spatial coordinates. 
Thus, we have 
\begin{eqnarray*}
a_{2n} (D^2) &=& \int\tilde{a}_{2n}(x, D^{2})\,dvol_g 
\\
&=& \textnormal{Vol}\cdot \tilde{a}_{2n}(D^{2}) 
\\
&=& 16\pi^2w_1w_2w_3 \, \tilde{a}_{2n}(D^{2})
\\
&=& 
\frac{w_1w_2w_3}{2\pi^{n+1}}\int_{\mathbb{S}^{2n+1}}\textnormal{tr}(\sigma_{-2-2n}(\Delta^{-1}))\,d^{2n+1}\zeta.
\end{eqnarray*}

The equation \eqref{structureformula} allows us to write 
$\sigma_{-2-2n}(x, \xi(\zeta)) = (w_1w_2w_3)^{-3n}P_{2n}(\zeta)$,  
which yields 
\begin{eqnarray*}
a_{2n} (D^2) &=& 
\frac{(w_1w_2w_3)^{1-3n}}{2\pi^{n+1}} \int_{\mathbb{S}^{2n+1}}\textnormal{tr}\left ( P_{2n}(\zeta))(\Delta^{-1}) \right ) \, d^{2n+1}\zeta \\
&=& (w_1w_2w_3)^{1-3n}Q_{2n}\left(w_1, w_2, w_3, w_1', w_2', w_3', \dots, w_1^{(2n)},  w_2^{(2n)},  w_3^{(2n)} \right ), 
\end{eqnarray*}
Note that $\textnormal{tr} \left ( P_{2n}(\zeta) \right )$ is a polynomial in 
$\zeta_1, \zeta_2, ... , \zeta_{2n+2}$, with the coefficients in 
the algebra generated by the rational numbers, trigonometric 
functions of the spatial coordinates, and $w_{i}^{(p)}$ where $i \in \{1,2,3 \}$, $p \in \{0,1,\dots,2n \}$.

The integral of a monomial 
$m_{\alpha}(\zeta) = c_{\alpha} \,\zeta_{1}^{\alpha_1} \cdots \zeta_{2n+2}^{\alpha_{2n+2}}$ 
over  $\mathbb{S}^{2n+1}$ is either $0$, or can 
be written as
\[
\int_{\mathbb{S}^{2n+1}}m_{\alpha}(\zeta)\, d^{2n+1}\zeta 
= 
\frac{2c_{\alpha}\prod_j \Gamma(\frac{\alpha_j+1}{2})}{\Gamma(n+1+\frac{|\alpha|}{2})},
\]
if each $\alpha_j$ is an even non-negative integer. Also, recall that 
$\Gamma(\frac{n}{2}) = q\pi^{\frac{1}{2}}$ for some 
$q\in\mathbb{Q}$ when $n\in2\mathbb{N}+1$, and 
$\Gamma(\frac{n}{2}) \in \mathbb{Z}$ when $n\in2\mathbb{N}$. Therefore 
we have  
\[
\int_{\mathbb{S}^{2n+1}}m_{\alpha}(\zeta)\, d^{2n+1}\zeta 
= 
q\pi^{\frac{2n+2}{2}} = q\pi^{n+1},
\] 
for some $q\in\mathbb{Q}$ if $c_{\alpha}\in\mathbb{Q}$.  Since 
$a_{2n} (D^2) = (w_1w_2w_3)^{1-3n}Q_{2n}$ has no 
spatial dependence, 
we conclude that 
\[
Q_{2n} 
=
\frac{1}{2\pi^{n+1}}  \int_{\mathbb{S}^{2n+1}}\textnormal{tr}(P_{2n}(\zeta))(\Delta^{-1})) \, d^{2n+1}\zeta
\] 
belongs to the algebra generated by the $w_{i}^{(p)}$ and rational numbers.
\end{proof}
\end{theorem}

\section{Gravitational Instantons, Modular forms, and Rationality}
\label{ThetaSection}

Among the Euclidean Bianchi type-IX models, an especially interesting 
class consists of the Bianchi IX gravitational instantons. A 
gravitational instanton is both self-dual (that is, the Weyl curvature 
tensor is self-dual) and an Einstein metric (the Ricci tensor is proportional 
to the metric). A remarkable feature of Bianchi IX gravitational instantons
with $SU(2)$ symmetry is that they can be completely classified in terms 
of solutions to Painlev\'e VI integrable systems, \cite{Hitchin, Oku, Tod}.  
The latter are a $4$-parameter family of singular ordinary differential 
equations of the form
\begin{eqnarray*}
\frac{d^2X}{dt^2}&=&\frac{1}{2}\left(
\frac{1}{X}+\frac{1}{X-1}+\frac{1}{X-t}\right)
\left(\frac{dX}{dt}\right)^2  
-\left( \frac{1}{t}+\frac{1}{t-1}+\frac{1}{X-t}\right)\frac{dX}{dt} \\&&
+\frac{X(X-1)(X-t)}{t^2(t-1)^2}
\left(\alpha +
\beta\frac{t}{X^2}+\gamma\frac{t-1}{(X-1)^2}+
\delta\frac{t(t-1)}{(X-t)^2}\right). 
\end{eqnarray*}

The self-dual equation for the $SU(2)$ Bianchi IX 
metrics is written in \cite{Oku} as an ordinary differential 
equation in the $w_i$ and in additional functions $\alpha_i$, 
$i=1,2,3$ that arise as the components of the
connection $1$-form in a basis of anti-self-dual $2$-forms, see \cite{Tod}.  
In terms of the conformally invariant variable 
$x = (\alpha_2-\alpha_1)(\alpha_2-\alpha_3)^{-1}$
the self-dual equations for the Riemannian Bianchi IX metric can be rephrased as
a system of equations
$$ w_i = \Omega_i x^\prime (x (1-x))^{-1/2}, $$
$$ \Omega^\prime_1 = -\frac{\Omega_2 \Omega_3}{x (1-x)}, \ \ \  \Omega_2^\prime = - \frac{\Omega_3 \Omega_1}{x}, \ \ \ 
\Omega_3^\prime = -\frac{\Omega_1 \Omega_2}{1-x}. $$
These in turn can then be reduced to a case of the Poincar\'e VI equation with parameters 
$$(\alpha ,\beta ,\gamma , \delta)=(\frac{1}{8},  -\frac{1}{8}, \frac{1}{8},  \frac{3}{8}),$$
see \cite{Oku, Tod}. 
In \cite{BaKo}, the solutions to this equation are given explicitly in terms of a parameterization involving
theta functions and theta characteristics
$$
\vartheta [p,q](z, i\mu ):=\sum_{m\in {\mathbb Z}} \exp \left( -\pi  (m+p)^2\mu
+ 2\pi i (m+p)(z+q)\right).
$$
Namely, with the notation $\vartheta [p,q]:= \vartheta [p,q] (0,i\mu )$, and 
$$ \vartheta_2:= \vartheta [1/2,0], \ \ \ \ \ 
 \vartheta_3:= \vartheta [0,0], \ \ \ \  \   \vartheta_4:=  \vartheta [0,1/2], $$
 one finds $\alpha_i = 2\, \partial_\mu \log \vartheta_{i+1}$ and
$$
w_1 =-\frac{i}{2}\vartheta_3\vartheta_4 \frac{\frac{\partial}{\partial q} \vartheta [p,q+\frac{1}{2}]}{e^{\pi ip} \vartheta [p,q]},\ \ \ 
w_2 =\frac{i}{2}\vartheta_2\vartheta_4 \frac{\frac{\partial}{\partial q} \vartheta [p+\frac{1}{2},q+\frac{1}{2}]}{e^{\pi ip} \vartheta [p,q]},
$$
$$
w_3 =-\frac{1}{2}\vartheta_2\vartheta_3 \frac{\frac{\partial}{\partial q} \vartheta [p+\frac{1}{2},q]}{ \vartheta [p,q]}.
$$
The asymptotics of these solutions were analyzed in \cite{ManMar}, where it is shown that, for large $\mu$,
they approximate Eguchi-Hanson type gravitational instantons with $w_2 = w_3 \neq w_1$, \cite{EgHa}.

It is clear that, for the Bianchi IX gravitational instantons, using the parameterization of \cite{BaKo},
the Seeley-de Witt coefficients $a_{2n}$ of the spectral action are rational functions, with 
${\mathbb Q}$-coefficients, in the $\vartheta_2, \vartheta_3, \vartheta_4$,  $\vartheta[p,q]$, 
$\partial_q \vartheta[p,q]$ and $e^{i\pi p}$ and derivatives, hence they belong to the field generated, over ${\mathbb Q}$,
by these functions. We will return in a second part of this work \cite{FFM2} to discuss in detail the arithmetic
properties of the spectral action for Bianchi IX gravitational instantons.

In this perspective, one can view the rationality question about the spectral action in a similar
light to analogous questions that occur whenever arithmetic and number theoretic structures arise
in theoretical physics. For example, when Feynman integrals are interpreted as periods
(see \cite{MarBook} for an overview of that setting), the fact that the relevant amplitude forms
and domains of integration are algebraic over ${\mathbb Q}$ (or ${\mathbb Z}$) has direct
implications on the class of numbers that arise as periods. Another such instance of
arithmetic structures in physics, where rational coefficients play an important role, is in the
zero temperature KMS states of quantum statistical mechanical systems: in the case 
constructed in \cite{CoMa} (see also Chapter 3 of \cite{ConMarBook}) for instance, the
construction of an arithmetic algebra of observables, defined over ${\mathbb Q}$, is linked 
to modular functions and makes it possible to have KMS states with values in the modular field. 
The relation between the spectral action of Bianchi IX gravitational instantons and modular forms
will be discussed in \cite{FFM2}.

\section{Conclusions} \label{ConclusionsSection}

We have shown that the Seeley-de Witt coefficients $a_{2n}(D^2)$
associated with the Dirac operator $D$ of the Bianchi type-IX 
metric, which appear in the expansion of the spectral action 
\cite{ChaConSAP},  are expressed by polynomials with rational 
coefficients evaluated on the cosmic evolution factors 
$w_1(t), w_2(t), w_3(t)$, and their derivatives of certain orders. 
It is quite interesting that although this metric provides a homogeneous 
anisotropic cosmological model, after remarkable cancellations, only 
rational coefficients appear in the final expression for each  
$a_{2n}(D^2)$. Such a rationality result was first conjectured in \cite{ChaConRW} 
for Robertson-Walker metrics, which was addressed in \cite{FatGhoKha}.

Our proof of the rationality statement for the Bianchi type-IX model, 
similar to the argument given in \cite{FatGhoKha}, begins with the 
crucial observation that the kernel of $e^{-tD^2}$  
is restricted to have no spatial dependence on the diagonal. We then 
take a novel approach to proceed the argument. That is, we 
have devised a general method  that expresses the Seeley-de Witt 
coefficients of a geometry as noncommutative residues of operators.  
This is an efficient method that allows explicit calculations with significantly 
less complexities, compared to the method of using parametric 
pseudodifferential calculus \cite{GilBook1}. More importantly, it leads to an 
elegant proof of the rationality result for the Bianchi type-IX metric. To be 
more explicit,  the Wodzicki residue \cite{Wod, Wod2} involves an integration over the 
cosphere bundle of a manifold, and the expression for computing 
$a_{2n}(D^2)$ simplifies to our favor when restricted to the cosphere 
bundle, in the view of our method.

\section*{Acknowledgments}

The first author is supported by
a Summer Undergraduate Research Fellowship at Caltech.
The third author is partially supported by NSF grants
DMS-1201512 and PHY-1205440 and by the Perimeter
Institute for Theoretical Physics.

\end{document}